\documentstyle[twocolumn]{article}
\renewcommand{\ref}[1]{\raisebox{.6ex}{[#1]}}

\newcommand{\be}{\begin{equation}}
\newcommand{\ee}{\end{equation}}

\begin{document}


\onecolumn

\footnotesize
\noindent
{\it Journal of Superconductivity, Vol. 8, No. 1, 1995 }

\vspace{0.80in}

\LARGE

\noindent
{\bf A Scenario to the Anomalous Hall Effect
        in the Mixed State of Superconductors}

\large

{\ }

\noindent
{\bf Ping Ao$^1$ }

{\ }


\small

\begin{flushright}
\begin{tabular}{p{5.in}}
\footnotesize
{\it Received 4 January 1995 }
\small
\vspace{0.05in}\\
\hline
\vspace{0.03in}
We argue that the motion of vacancies in a pinned vortex lattice may
dominate the contribution to the Hall effect in an appropriate parameter
regime for a superconductor.
Based on this consideration
a model is constructed to explain the anomalous Hall effect
without any modification of the basic vortex dynamic equation.
Quantitative predictions are obtained.
Present model can be directly tested by an observation of the vacancy motion.
\vspace{0.05in}\\
\hline
\vspace{0.03in}
\end{tabular}
\end{flushright}

\normalsize

\clearpage

\twocolumn

It is generally believed that the motion of vortices in a
superconductor is responsible for the apparence of resistance.
Indeed, there is a well and successful
account of dissipative phenomena in superconductors
based on this idea\cite{parks}.
However,
the ubiquitous occurrence of the anomalous Hall effect in the mixed state of
both conventional and oxide superconductors, the sign change of the
Hall resistance below the superconducting transition temperature,
seems to be inexplainable by standard models for vortex dynamics
in superconductors\cite{exp}.
The purpose of the present paper is to show that
it is nevertheless  possible to understand the anomalous Hall effect without
either a modification of vortex dynamics
or assuming two types of carriers(or more complicated electronic
band structures).
The essential idea in the present model is the domination
of the vacancy motion
in a pinned vortex lattice under a suitable condition.
Contrast to previous ones in this model
both the strong interaction between vortices, which leads to the
many-body correlation effect, and the strong pinning effect are crucial.
In the following we present the arguments leading to the model,
and discuss its predictions and experimental verifications.
For simplicity, we will consider an isotropic s-pairing superconductor with
one type of charge carriers in two dimension.
In this situation vortices(or straight vortex lines)
can be viewed as point particles.

{\ }

\footnotesize

\noindent
\begin{tabular}{p{3.17in}}
\hline
\vspace{0.01in}
$^{1}$Department of Theoretical Physics,
Ume\aa{\ }University, S-901 87, Ume\aa, SWEDEN
\end{tabular}

{\ }

\normalsize

\newpage

We begin with the discussion of vortex dynamics.
The equation of motion for a vortex generally takes the form of the
Langevin equation similar to that of a
charged particle in the presence of a magnetic field:
\be
   m_v \ddot{\bf r} = q_v \frac{\rho_s}{2} h d \; ( {\bf v}_s - \dot{\bf r} )
    \times \hat{z} - \eta \dot{\bf r} + {\bf F}_{pin} + {\bf f} \; ,
\ee
with an effective mass $m_v$,
a pinning force ${\bf F}_{pin}$, a vortex viscosity $\eta$, and a
fluctuating force ${\bf f}$. The viscosity is related to the fluctuating
force by the usual fluctuation-dissipation theorem.
In Eq.(1) $q_v = \pm 1$ describes the vorticity, $h$ is the Planck
constant, $\rho_s$  the superfluid electron number density at temperature $T$,
$d$ the thickness of the superconductor film, and $\hat{z}$ the unit
vector in $z$-direction.
Normal current is assumed to be zero and
a uniform temperature in the whole film is also assumed in Eq.(1).
The term associated with the superconducting electron velocity ${\bf v}_s$
and the vortex velocity $\dot{\bf r}$ at the right side of Eq.(1)
is the so-called Magnus force.
It has been shown that under the general properties of
a superconductor it must take such a form\cite{ao1}.
Closely related but different phenomenological equations
for vortices have been discussed in the literature\cite{bardeen}.
Eq.(1) is the starting point in recent studies of vortex tunneling in
superconductors\cite{ao2}.

We first consider the measured electric field generated by a steady motion of
vortices and the resulting Hall effect.
In a steady state, if there is no pinning force, i.e. ${\bf F}_{pin}= 0$,
from Eq.(1) the velocity of the vortices is

\newpage

\be
   \dot{\bf r} = \frac{ (\rho_s d h/2)^2}{ \eta^2 + (\rho_s d h/2)^2 } \;
                 {\bf v}_s
          + q_v \; \frac{(\rho_s d h/2) \eta }{ \eta^2 + (\rho_s d h/2)^2 } \;
                 {\bf v}_s \times\hat{z} \; .
\ee
According to the Josephson relation, the motion of vortices produces
phase slippages, therefore generates an electrochemical
potential difference in a superconductor,
which is the reaction force to the superconducting electrons
due to the vortex velocity part of the Magnus force in Eq.(1).
This potential difference
corresponds to the measured electric field ${\bf E}$. For a vortex density
$n$ with an average velocity $\dot{\bf r} = {\bf v}_l$, we have
\be
   {\bf E } = - q_v \frac{h}{2e} \; n \; {\bf v}_l \times \hat{z}  \; .
\ee
If the density of moving vortices is equal to the vortex density induced
by the applied perpendicular magnetic field ${\bf B} = B \hat{z} $,
$n = |B|/\Phi_0$ with $\Phi_0 = h c/2|e|$ the flux quantum of Cooper pairs,
we have the usual expression
\be
      {\bf E } = - \frac{1}{c}   {\bf v}_l \times {\bf B}  \; .
\ee
Therefore, in the steady state and in the absence of pinning,
all vortices move with the same velocity
and we have the
longitudinal resistivity $\rho_{xx}$ as
\[
   \rho_{xx} = \frac{E_x}{J}
             = \frac{q_v}{ec}\; \frac{(d h/2) \eta}{\eta^2 +
                  (\rho_s d h/2)^2 }  \; B \;
\]
\be
             = \frac{\Phi_0}{c^2}\; \frac{ \eta d }{\eta^2 + (\rho_s d h/2)^2 }
                \; |B| \;  ,
\ee
and the transversal(Hall) resistivity $\rho_{yx}$ as
\[
   \rho_{yx} = \frac{E_y}{J} = \frac{1}{ec}\; \frac{(d h/2)^2 \rho_s }
                              {\eta^2 + (\rho_s d h/2)^2 } \; B \;
\]
\be
              = q_v \frac{\Phi_0}{c^2 }\; \frac{\rho_s d^2 h/2 }
                              {\eta^2 + (\rho_s d h/2)^2 }   \; |B| \; .
\ee
Here the current density ${\bf J} = e \rho_s {\bf v}_s$ is along
$x$-direction.
The relation between vorticity $q_v$, the charge sign of carriers, and the
direction of the magnetic field is $q_v {e}/{|e|} = {B}/{|B|}$.
It is clear that from Eq.(6) the Hall effect in the superconducting state has
the same sign as that of the normal state.
This, and also those similar conclusions based on
the classical models for the vortex dynamics in superconductors,
are in apparent contradiction with many experimental observations\cite{exp},
the so-called anomalous Hall effect.

There is a serious problem related to
the many-body correlation between vortices and the pinning effect
in the using of the above simple minded result, Eq.(6),
for real superconducting materials.
The many-body correlation due to the strong and long range interaction between
vortices gives arise to the Abrikosov lattice in the mixed state of
a superconductor.
However, even after the consideration of the strong many-body effect,
if there were no pinning for vortices,
the whole vortex lattice would move together under the influence of
an externally applied current in the same manner as that of
independent vortices. Hence one would get the same sign of the Hall
effect in both superconducting and normal states.
This will be changed in the presence of pinnings.
If we introduce some strong
pinning centers into the film and consider the edge pinning,
the vortex lattice will be pinned down.
In such a situation the motion of the vortex lattice is made possible
by various kinds of thermal fluctuations.
We will argue below that at low temperatures
the dominant contribution to the motion is due to vacancies.
We will advance our argument in the following way. First we will
show that the creations and motions of vacancies and
interstitials are energetically favored over vortices hopping into and out of
pinning centers as well as other motions.
Then we will argue that vacancies are even more
favourable than interstitials.

For two vortices separated by a distance $r$,
which is less than the effective magnetic
screening length $\lambda_{\perp}= \lambda_L^2 /d$($d< \lambda_{L}$,
$\lambda_{\perp}= \lambda_L$ if $d> \lambda_L$) but greater than $\xi_0$,
the interaction potential is\cite{gennes}
\[
   V_l(r) = 2 \left( \frac{\Phi_0}{4\pi \lambda_L } \right)^2 d \;
         \ln \frac{r}{\xi_0}
       = 2 \frac{ h^2 \rho_s }{8 \pi m^{\ast} } d \; \ln \frac{r}{\xi_0}
\]
\be
       \equiv 2 \epsilon_0   \ln \frac{r}{\xi_0} \; .
\ee
Here $\lambda_L^2 = m^{\ast} c^2/8 \pi \rho_s e^2  $ is
the London penetration depth,
$m^{\ast}$ the effective mass of a Cooper pair,
and $\xi_0$ the coherence length of the superconductor.
The energy scale
$\epsilon_0$ sets both the scale for the strength of vortex
interaction and the scale for the strength of a strong pinning center.
The interaction potential between a dislocation pair separated by a distance
larger than the lattice constant $a_0$ is given by\cite{brandt}
\be
   V_d(r) = \frac{1}{ 2\sqrt{3} \pi } \epsilon_0 \ln\frac{r}{a_0}
          \equiv 2 \epsilon  \ln\frac{r}{a_0}  \; .
\ee
The energy scale $\epsilon$ for the dislocation interaction here
is about 20 times smaller than
$\epsilon_0$ for the vortex interaction and pinning centers.
It is therefore energetically much more favorable to have dislocation pairs
in the lattice. Hence for temperature $T << \epsilon_0$ we can ignore the
contribution from the vortices hopping out of pinning centers(and also the
creation of vortex-antivortex  pairs).
The vortex lattice is then effectively pinned at such temperatures.
The precise calculation of the vacancy $\epsilon_v$ and interstitial energy
$\epsilon_i$ is difficult,
because the detailed information on the scale of lattice constant is needed.
However,
because vacancies and interstitials can be viewed as the smallest dislocation
pairs\cite{friedel},
we immediately have the estimated energy scale for $\epsilon_{v}$ as,
by putting $r \sim 2 a_0$ in Eq.(8),
\be
    \epsilon_v \sim 2 \epsilon  =
           \frac{1}{ 2\sqrt{3} \pi }
            \left(\frac{ \Phi_0 }{4\pi \lambda_L} \right)^2 d \; .
\ee
There might be a weak dependence on the magnetic field through the vortex
lattice constant, which is neglected here.
It is clear from the above analysis
that vacancies and interstitials have the lowest excitation energy scale.

Now let us compare the creation energy of a vacancy
and that of an interstitial.
At low magnetic fields
the vacancy formation energy is lower than that of interstitials,
because of the strong short range and repulsive nature of the
interaction between vortices.
This has been shown theoretically\cite{hill}.
Therefore vacancies will dominate over interstitials at low temperatures and
low fields, which has been observed experimentally\cite{energy}.
Similar phenomenon has also been observed in other crystalline
structures\cite{friedel}.
At high magnetic fields the vortex interaction becomes of long range
comparing with the vortex lattice constant.
The short range behavior which determines the formation energy difference
between vacancies  and interstitials becomes relatively unimportant.
Although one expects a zero formation energy difference for
an infinite range interaction, vacancies may still have a lower formation
energy for a finite range interaction.
This is based on the plausible expectation that the formation
energy difference between vacancies and interstitials
increases monotonically to zero with the interaction range.
We will not be surprised if the anomalous Hall effect is more
pronounced at lower magnetic fields.

We examine its experimental consequencies if the vacancy motion dominates.
Effectively, the motion of a vacancy in the pinned
vortex lattice behaves as an antivortex, that is, with a vorticity $-q_v$
under the action of an applied supercurrent.
This leads us to our main conclusion that at low enough
temperatures the sign of the Hall resistance is different from its sign in
the normal state.
Quantitatively, the motion of vacancies is governed by an equation of
the same form as Eq.(1),
and can be viewed as independent particles moving in the
periodic potential formed by the vortex lattice and a random potential
due to the residue effect of pinnings.
The potential height of the periodic potential as well as that of
the random potential is presumably the order
of $\epsilon_v$.
Assuming the vacancy density $n_v$ in a steady state,
following the same procedure as from Eq.(1) to
Eqs.(5-6), the longitudinal
resistivity
is then
\be
    \rho_{xx} = \frac{h}{2 e^2} \; \frac{\eta_{eff} \; \rho_s d h /2 }
                      { \eta_{eff}^2 + (\rho_s d h /2)^2 } \;
                \frac{ n_v }{\rho_s }  \; ,
\ee
and the Hall resistivity
\be
   \rho_{yx} = - q_v \frac{h}{2 e^2} \; \frac{(\rho_{s}dh/2)^2 }
                   { \eta_{eff}^2 + (\rho_s dh/2)^2 } \;
                \frac{ n_v }{\rho_s}  \; .
\ee
It is evident that the sign for $\rho_{yx}$ in Eq.(11) is opposite to that
in Eq.(6).
Here $\eta_{eff}$ is the effective vacancy viscosity,
related to the vacancy diffusion constant in the
periodic potential due to the vortex lattice by the Einstein relation between
the diffusion constant and the mobility.
We expect that $\eta_{eff} = \eta_0 \; e^{a \; \epsilon_v /K_B T } $
with $a$ a numerical factor of order unity at low temperatures.
In the low temperature limit, superfluid electron number density
$\rho_s$ approaches a constant, vacancy density $n_v= n_0 \;
e^{ - b \; \epsilon_v / k_B T}$ may be
exponentially small($b= 1 $ for the thermally activated vacancies, and
$b =0$ for the pinning center induced vacancies),
and the effective vacancy viscosity $\eta_{eff}$
can be exponentially large.
In this limit both longitudinal and Hall resistances vanish
exponentially, and
we obtain a scaling relation between
the Hall and longitudinal resistivities as
\be
   \rho_{yx} = A \; \rho_{xx}^{\nu } \; ,
\ee
with $A = - q_v (\rho_s dh/2\eta_0)^{b/(a + b)}
          (2e^2 \rho_s/h n_0 )^{a/(a + b)}  $
and the power $\nu = {(2 a + b)} /{(a + b)}$
varies between 1 and 2 depending on the detail of a sample which determines
the numerical factors $a$ and $b$.
It is interesting to note that the present range of $\nu$ covers
the values obtained previously by completely different approaches\cite{dorsey}.
We also note that if the vortex (vacancy) tunneling process\cite{ao2}
is allowed,
there will be a finite Hall resistance at zero temperature.

The linear film thickness $d$ dependence of the vacancy formation energy
$\epsilon_v$, Eq.(9), can also be tested experimentally.
We expect this linear dependence to hold for thin enough films.
For a thick film a vortex line will bend in the $z$-direction,
and the straight line assumption in reaching Eq.(9) is no longer valid.
This suggests a critical thickness $d_c$ such that
 Eq.(9) is valid for $d< d_c$ and $\epsilon_v$ will be saturated for $d>d_c$.
This critical thickness may reflect
the many-body correlation along the direction of the magnetic field
and is likely a sensitive function of both
the magnetic field and the temperature.

Now we discuss the qualitative features of the model.
In the above picture, we need the vortex lattice to define vacancies, and
a sufficiently strong
pinnings to prevent the slidding of vortex lattice in order to obtain
a maximum contribution of vacancies.
However, there is no need for a whole lattice structure.
Sufficiently large local structures, like lattice domains, will be enough to
define vacancies. Therefore one may even have vacancy like excitations in
a vortex liquid state, where large local orderings exist.
Whether or not this is also true for a vortex glass state depending on its
randomness.
On the other hand,
if the pinning is too strong, for example, the pinning center density
is much larger than the vortex density, vortices will
be individually pinned down and the local
lattice structure required
for the formations of  vacancies and interstitials will be lost.
Therefore we only expect to see the anomalous
Hall effect in a suitable range of pinnings and magnetic fields, that is,
for $B_l < |B| < B_u$ with the lower and upper critical fields
determined by pinnings.

As shown by Eqs.(7-9) a large London penetration
depth reduces down all the relevant energy scales,
and energetically it is more favorable to observe the anomalous Hall effect.
This may explain the experimental observation that it is difficult
to observe the anomalous Hall effect in a clean conventional
superconductor, but easy in a dirty superconductor as well as in a high
temperature superconductor where the London penetration depth is
large\cite{exp}.
At this point, we wish to point out the important difference between the
pinning centers for vortices and the mean free path of electrons:
The vortex dynamics is sensitive to pinning, not to the mean free path.
There is no obvious relationship between them.

If the temperature is increased towards the superconducting transition
temperature, all energy scales become small, and so does the difference
between $\epsilon_v$ and $\epsilon_i$.
Other contributions, such as
those of quasiparticles, the slidding of vortex lattice,
and vortices hopping out of pinning centers, become dominant.
Those contributions have a different sign for the Hall resistance,
but the same sign for the longitudinal resistance as vacancies.
Therefore one expects a sign change in the Hall resistance below
the transition temperature.

It is important to check whether or not the present model can also explain
experiments concerning vortex motion other than the
Hall and longitudinal resistivity experiments.
We will mention here the Nernst effect.
Under the driving of a temperature gradient the force felt by a vacancy
is opposite of the force felt by an interstitial or a vortex in direction
but equal in magnitude.
Then the Nernst effect due to vacancies has the same sign as that of
vortices or interstitials. Therefore our model gives that
in the anomalous Hall effect regime there is no sign change for the Nernst
effect, and furthermore, the Nernst effect is more pronounced because of the
large contribution due to both vacancies and interstitials.
This is in agreement with the experimental observations\cite{hagen}.

In conclusion,
a model based on the motion of vacancies in a pinned lattice
is proposed to explain the anomalous Hall effect.
No modification of the basic vortex dynamics equation is needed.
Quantitatively it leads to an exponential tail and the scaling relation
at low temperatures, and no sign change for the Nernst effect.
For thin enough films the activation energy in the low temperature limit
has a linear film thickness dependence.
Present model provides a framework to understand relevant experiments, and
further experiments are needed to test it.
The most direct way might be the observation of the vacancy motion.

{\ }

\noindent
{\bf ACKNOWLEDGEMENTS }

{\ }

  Valuable discussions with David Thouless and
  Andrei Shelankov, in particular their sharp remarks,
  as well as the correspondences on experimental aspects of the problem
  from John Graybeal and Chris Lobb,
  are gratefully acknowledged.

{\ }

\end{document}